\documentclass[conference]{IEEEtran}
\IEEEoverridecommandlockouts

\usepackage{cite}
 \usepackage{url}
\def\BibTeX{{\rm B\kern-.05em{\sc i\kern-.025em b}\kern-.08em
    T\kern-.1667em\lower.7ex\hbox{E}\kern-.125emX}}

\usepackage{graphicx}%
\usepackage{dblfloatfix}
\usepackage{multirow}%
\usepackage{amsmath,amssymb,amsfonts}%
\usepackage{amsthm}%
\usepackage{mathrsfs}%
\usepackage{xcolor}%
\usepackage{textcomp}%
\usepackage{manyfoot}%
\usepackage{booktabs}%
\usepackage{algorithm}%
\usepackage{algorithmicx}%
\usepackage{algpseudocode}%
\usepackage{listings}%
\usepackage{comment}
\usepackage{pifont}
\usepackage[section]{placeins}
\usepackage{amssymb}

\newcommand{\keepcomment}{1} 
\usepackage[normalem]{ulem}
\ifnum\keepcomment=1
	\usepackage[colorinlistoftodos,textsize=scriptsize]{todonotes} 
    \newcommand{\stkout}[1]{\ifmmode\text{\sout{\ensuremath{#1}}}\else\sout{#1}\fi}
    
\else
	
	\usepackage[disable]{todonotes} 
\fi

\usepackage{amsfonts} 

\usepackage{calrsfs} 
\DeclareMathAlphabet{\pazocal}{OMS}{zplm}{m}{n}

\usepackage[normalem]{ulem}   
\usepackage{xcolor}          
\usepackage{amsmath}         

\newif\ifshowchanges
\showchangesfalse  

\newcommand{\deleted}[1]{%
  \ifshowchanges
    \ifmmode
      \text{\textcolor{red}{\sout{$#1$}}} 
    \else
      \textcolor{red}{\sout{#1}}        
    \fi
  \fi
}

\usepackage{mathtools} 
\usepackage{calrsfs} 
\usepackage{float}
\usepackage{overpic}   
\usepackage{booktabs} 
\usepackage{subcaption} 
\usepackage{placeins}

\theoremstyle{thmstyleone}%
\newtheorem{theorem}{Theorem}

\newtheorem{proposition}[theorem]{Proposition}

\newtheorem{definition}[theorem]{Definition}
\newtheorem{assumption}[theorem]{Assumption}

\newcommand\defeq\triangleq
\newcommand\defequiv\coloneqq

\author{
    \IEEEauthorblockN{
        Amal Sakr$^{1}$, Andrea Araldo$^{1}$, Tijani Chahed$^{1}$, Daniel Kofman$^{2}$
    }
    \IEEEauthorblockA{
        $^{1}$SAMOVAR, Télécom SudParis, Institut Polytechnique de Paris, 91120 Palaiseau, France \\
$^{2}$Télécom Paris, Institut Polytechnique de Paris, 91120 Palaiseau, France \\
        \{amal.sakr, andrea.araldo, tijani.chahed\}@telecom-sudparis.eu, \\
        daniel.kofman@telecom-paristech.fr
    }
}
\usepackage[colorlinks=true,
            linkcolor=blue,
            citecolor=blue,
            urlcolor=blue]{hyperref}
\begin{document}

\title{Co-Investment in Mobile Edge Computing 
with Infrastructure Update and Dynamic Participation}

\maketitle
\begin{abstract}
Mobile Edge Computing (MEC) requires Network Operators (NOs) to undertake substantial infrastructure investments, while most revenues are captured by Service Providers (SPs) offering end-user applications. This cost-revenue imbalance discourages NOs from investing in MEC deployment, despite increasing demand for low-latency and bandwidth-intensive services.
This paper proposes a co-investment scheme in which players, i.e., one NO and multiple SPs, jointly deploy, maintain, and share MEC infrastructure over multiple decision epochs. We devise a new coalitional game model that captures the planning of resources, their allocation among players, and cost and revenue sharing. To address fluctuating user demand and evolving participation incentives, we design a mechanism that updates resources and allows the dynamic entrance and exit of players over time. We sustain cooperation through a compensation scheme. Numerical results show that combining resource updates with dynamic participation increases the total payoff and strengthens the NO’s incentive to invest.
\end{abstract}


\section{Introduction}

Mobile Edge Computing (MEC) depends on distributed infrastructure deployed close to end users in order to support general-purpose, low-latency and bandwidth-intensive compute services~\cite{tran2017collaborative}, open to end-users of different Service Providers (SPs). This requires Network Operators (NOs), such as Comcast, to bear substantial capital expenditures (CAPEX) and operational expenditures (OPEX) for deploying and maintaining edge resources~\cite{tun2019energy}. However, NOs argue that these investments may not generate a sufficient return~\cite{GSMA2023FairShare}. This concern is reinforced by the current revenue structure: third-party SPs, such as Netflix or future virtual-reality service providers, capture most revenues from end users, while the underlying infrastructure costs are mainly sustained by NOs~\cite{Telefonica2022OTTtraffic}. Hence, large-scale MEC deployment requires a fairer allocation of investment responsibilities, where SPs also contribute to the infrastructure from which they benefit~\cite{GSMA2023FairShare}. Yet, despite the technological readiness of MEC, no widely adopted investment model has emerged, and MEC remains under-deployed~\cite{ieee2022mec}.

To address this barrier, we propose a co-investment scheme in which players, namely a NO and multiple SPs, jointly deploy and maintain MEC infrastructure, while sharing both costs and revenues. This setting involves strategic interactions among heterogeneous players with different roles and incentives. In particular, the NO provides the infrastructure, whereas SPs exploit the deployed resources to serve end users. As a result, individual incentives may not align with collective efficiency, since players may benefit differently from the same infrastructure. We therefore model MEC co-investment as a coalitional game, which allows us to capture how groups of players cooperate, how the joint payoff is generated, and how this payoff can be allocated among participants. 

In addition, user demand for MEC services may vary over time due to daily traffic patterns, seasonal effects, and changes in service popularity. Hence, deciding how much edge capacity to deploy and how to allocate it among SPs is a key challenge. To address this issue, we allow infrastructure capacity (e.g., resources such as CPU) and resource allocation to be updated over time. This enables the coalition to better match deployed resources with traffic variations, reducing both unused capacity and capacity shortages. A further challenge is that player participation in MEC co-investment may evolve over time. 
Players may join, remain in, or leave the co-investment depending on the payoff they obtain from cooperation. 
While dynamic participation improves flexibility, it may reduce the payoff of players who remain in the coalition. Therefore, a mechanism is needed to regulate coalition changes and compensate affected players.

Our main contributions are:
\begin{itemize}
    \item We model a MEC co-investment scheme between a NO and SPs as a coalitional game, in which players jointly deploy and maintain edge infrastructure, share costs and revenues, and make network planning decisions (edge capacity and allocation) to maximize payoff.
    \item We design a mechanism to regulate the updates of the infrastructure resources, and the decisions of players to join, leave, or remain in the co-investment. Such actions are subject to monetary compensations, which we calculate so as to encourage cooperation over time.
\end{itemize}

We show numerically that infrastructure updates and dynamic participation over time improve MEC co-investment performance. Together, they increase both the total payoff and the NO’s payoff. Thus, the proposed co-investment scheme strengthens the NO’s incentive to deploy new MEC infrastructure and sustain long-term cooperation with SPs.
We release our code as open source to ensure reproducibility~\cite{ourcode}.
\clearpage
The remainder of the paper is organized as follows. Section~\ref{sec:related} reviews related work. Section~\ref{sec:model} presents the system model and coalitional game formulation. Section~\ref{sec:dynamic} introduces the dynamic participation mechanism. Section~\ref{sec:res} discusses the numerical results. Section~\ref{sec:conclusion} concludes the paper.

\section{Related Work}\label{sec:related}

Cooperation between NOs and SPs already exists in practice. For instance, Netflix engages in paid peering with NOs and deploys caching infrastructure within their networks, reflecting technical and commercial cooperation~\cite{NetflixOpenConnect}. Similar arrangements also arise when SPs are bundled with NOs, including revenue-sharing mechanisms~\cite{limbach2014cooperative}. Our work is grounded in these existing forms of cooperation and extends them by enabling joint investment in network infrastructure.

Existing MEC studies mainly focus on resource allocation, pricing, or utilization over already deployed infrastructure. Examples include resource slicing by NOs~\cite{tun2019energy}, cooperation among edge nodes~\cite{tran2017collaborative}, joint resource planning over existing edge nodes~\cite{xiang2023game}, multi-stage task offloading and resource allocation~\cite{zhang2021deployment}, resource rental by SPs~\cite{li2020data}, and infrastructure utilization improvement~\cite{kim2019economics}. In contrast, our work focuses on cooperative investment between a NO and SPs for jointly deploying and maintaining new MEC infrastructure.

Cooperative infrastructure deployment has also been studied through co-investment~\cite{he2024co}, joint budget allocation~\cite{koutsopoulos2019economics}, and bilateral investment models~\cite{INDERST201428,bourreau2021co}. However, these works typically consider homogeneous cooperating entities. In contrast, we consider heterogeneous players with different roles and incentives: the NO owns and operates the infrastructure, while SPs generate revenue by using it. 

Closest to our work,~\cite{sakr2025co,sakr2025coj} study MEC co-investment between a NO and SPs, but assume that both the deployed capacity and the set of participating players are fixed in advance. In contrast, we consider a multi-epoch setting in which infrastructure capacity can be updated to follow demand variations, and participation decisions can evolve over time. 

Dynamic coalition formation has been studied in different contexts. Sequential coalition formation is considered in~\cite{bloch1996sequential}, where players join coalitions through proposals but cannot revise their participation. Coalitions evolving over discrete time are analyzed in~\cite{lehrer2013core}, where deviations lead to subgames without re-merging. 

Other works consider adaptive coalition changes, such as payoff-based participation updates~\cite{peleteiro2014fostering}, merge-and-split rules~\cite{zhao2018edge}, and join-leave-remain decisions~\cite{chituc2007join}. However, these works do not jointly address infrastructure dimensioning, payoff allocation, and the use of transfer mechanisms to regulate participation and support cooperation over time.

To the best of our knowledge, this is the first work to model MEC co-investment with both infrastructure updates and dynamic participation, while using transfer fees, exit penalties, and compensations to regulate coalition transitions.
\newpage
\section{System Model}\label{sec:model}

We consider a Mobile Edge Computing (MEC) system where a set of players 
\( \pazocal{N} \coloneqq \{NO, SP_1, \dots, SP_N\} \) 
consists of one Network Operator (NO) and multiple Service Providers (SPs).
The investment period $I$ is divided into decision intervals~\( \pazocal{I}_k \), indexed by decision epochs 
\( k \in \pazocal{T} \coloneqq \{1,\dots,K\} \).
Each interval \( \pazocal{I}_k \) has duration \( \Delta \) (e.g., one year).

At the beginning of each decision epoch \(k\), 
players decide whether to participate in the co-investment, forming a coalition 
\( \pazocal{S}_k \subseteq \pazocal{N} \).
If a coalition \( \pazocal{S}_k \) forms, its participants jointly decide the total MEC capacity 
\( C_k \in \mathbb{R}_+ \) to deploy. This capacity is then maintained over the corresponding decision interval \( \pazocal{I}_k \).
The deployment and maintenance cost, denoted by 
\( \text{Cost}(C_k, C_{k-1}) \), depends on the previously installed capacity \(C_{k-1}\) and the new capacity \(C_k\).
We set \( C_0 = 0 \), so that \( \text{Cost}(C_1, C_0) \) corresponds to the initial deployment cost.

The cost includes (i) a capacity adjustment cost, capturing infrastructure upgrades or reductions, including fixed costs associated with technical intervention, and (ii) a maintenance cost proportional to the deployed capacity $C_k$ and the duration of the decision interval $\pazocal{I}_k$.

Each decision interval $\pazocal{I}_k$ is divided into time slots \( t \). 
At each time slot, the NO allocates the available capacity \( C_k \) among the SPs in \( \pazocal{S}_k \).
Let \( h_{i,k}^t \in \mathbb{R}_+ \) denote the amount of capacity allocated to SP \(i \in \pazocal{S}_k \) at time slot \(t\), subject to:
\begin{equation}
\sum_{i \in \pazocal{S}_k} h_{i,k}^t \le C_k, \quad \forall t \in \pazocal{I}_k
\end{equation}
Each SP uses its allocated capacity to serve users and derives utility 
\( u_{i,k}^t(h_{i,k}^t) \), representing its revenue at time slot~\(t\).

The total payoff of coalition \( \pazocal{S}_k \) at epoch \(k\) is given by:
\begin{equation}
v_k(\pazocal{S}_k) = \sum_{t} \sum_{i \in \pazocal{S}_k} u_{i,k}^t(h_{i,k}^t) 
- \text{Cost}(C_k, C_{k-1})
\label{eq:v_k}
\end{equation}

\subsection{Assumptions}

We make the following standard assumptions.

\begin{assumption}
\label{ass:utility}
For all \( i \in \pazocal{S}_k \) and time slots \(t \in \pazocal{I}_k\), 
the utility function \( u_{i,k}^t(h_{i,k}^t) \) is non-negative, non-decreasing, and concave, 
capturing diminishing returns. Moreover, no revenue is generated without allocated capacity, i.e., 
\( u_{i,k}^t(0)=0 \).
\end{assumption}
Non-negativity ensures that MEC usage does not generate negative revenue. 
Monotonicity reflects that allocating more capacity enables an SP to serve more requests or improve service quality, thereby increasing revenue.
Concavity captures diminishing marginal returns: after a certain level of allocation, 
additional capacity provides progressively smaller gains. 

\begin{assumption}
\label{ass:NO}
The NO does not directly generate revenue from MEC services, as it does not interact directly with end users and is not allocated any capacity. Hence, its utility is zero:
\(
u_{NO,k}^t(0) = 0, \quad \forall t.
\)
\end{assumption}
The NO derives its payoff as a share of the total payoff.

Given a coalition \( \pazocal{S}_k \), the capacity \( C_k \) and the allocation 
\( H_k = \{ h_{i,k}^t \}_{i \in \pazocal{S}_k,\, t \in \pazocal{I}_k} \)
are chosen to maximize the total coalition payoff:
\begin{equation}
\begin{aligned}
(C_k^*, H_k^*) := \operatorname*{argmax}_{C_k, H_k} \quad & v_k(\pazocal{S}_k) \\
\text{s.t.} \quad 
& \sum_{i \in \pazocal{S}_k} h_{i,k}^t \le C_k, \quad \forall t \in \pazocal{I}_k, \\
& h_{i,k}^t \ge 0, \quad \forall i \in \pazocal{S}_k,\; t \in \pazocal{I}_k
\end{aligned}
\label{eq:opt}
\end{equation}
where $v_k(\pazocal{S}_k)$ is defined in~\eqref{eq:v_k}.

\subsection{Coalitional Game Formulation}
To model MEC co-investment, we formulate the interaction among players 
as a coalitional game with transferable utility.
\begin{definition}
\label{def:game}
A coalitional game with transferable utility, capacity update and dynamic participation is defined in characteristic form over a finite set of decision epochs \( \pazocal{T} \coloneqq \{1, \dots, K\} \) as the pair
\(
\pazocal{G} = \left\langle \pazocal{N}, \{ v_k \}_{k \in \pazocal{T}} \right\rangle,
\)
where \( \pazocal{N} \) is the set of players and 
\( v_k: 2^{\pazocal{N}} \to \mathbb{R} \) assigns to each coalition 
\( \pazocal{S}_k \subseteq \pazocal{N} \) its value at epoch \(k\).
\end{definition}
The coalitional game in Definition~\ref{def:game} is not a sequence of independent games. Consecutive epochs are linked through past infrastructure capacity decisions and payoffs.

\subsubsection{Stability Notions}

A key question is whether a coalition remains stable over the upcoming decision interval $\pazocal{I}_k$, 
i.e., whether no subset of players has an incentive to deviate during that interval.

\begin{definition}
\label{def:stable}
A coalition \( \pazocal{S}_k \) is stable if there exists a payoff allocation 
\( \{x_{i,k}^{\pazocal{S}_k}\}_{i \in \pazocal{S}_k} \) such that:
\begin{enumerate}
    \item \textbf{Coalitional rationality (CR):}
    \[
    \sum_{i \in \pazocal{S}'} x_{i,k}^{\pazocal{S}_k} \ge v_k(\pazocal{S}'), 
    \quad \forall \pazocal{S}' \subseteq \pazocal{S}_k
    \]
    \item \textbf{Group rationality (GR):}
    \[
    \sum_{i \in \pazocal{S}_k} x_{i,k}^{\pazocal{S}_k} = v_k(\pazocal{S}_k)
    \]
\end{enumerate}
\end{definition}

\begin{proposition}
\label{prop:stab}
Fix epoch $k$. The coalition \( \pazocal{S}_k \subseteq \pazocal{N} \) is stable over interval $\pazocal{I}_k$, i.e., 
 there exists a stable payoff allocation.
\end{proposition}

\begin{proof}[Proof Sketch]
The result follows from the fact that the coalitional game 
\( (\pazocal{S}_k, v_k) \) has a non-empty core, the latter defined as a payoff allocation that satisfies Definition \ref{def:stable}. One such solution, albeit radical, is when total payoff is given to the NO, the SPs in this case will have payoff equal to zero, which does not prevent them from being in the coalition. This allocation satisfies Definition \ref{def:stable}, and hence the non-emptiness of the core, and the stability of the coalition.
\end{proof}

\section{Mechanism for Dynamic Participation}\label{sec:dynamic}

\subsection{Mechanism Design}

At epoch \(k\), players decide whether to enter, remain in, or leave the coalition.
These decisions are driven by the payoff they expect to obtain within the coalition.

Let \( \pazocal{S}_k \) and \( \pazocal{S}_{k-1} \) denote the coalitions at epochs 
\(k\) and \(k-1\), respectively.
To maintain coalition cohesion under such changes, we introduce transfer mechanisms.
A player \( i \in \pazocal{S}_k \setminus \pazocal{S}_{k-1} \) entering the coalition 
pays an entry fee \( f_{i,k} \), while a player 
\( i \in \pazocal{S}_{k-1} \setminus \pazocal{S}_k \) leaving the coalition 
incurs an exit penalty \( p_{i,k} \).
Persistent players who remain in the coalition, i.e., 
\( i \in \pazocal{S}_k \cap \pazocal{S}_{k-1} \), receive a compensation \( c_{i,k} \) 
to offset possible losses due to coalition reconfiguration.
At the initial epoch, no transfers apply:
\( f_{i,1} = p_{i,1} = c_{i,1} = 0 \) for all \( i \in \pazocal{N} \).

Compensation ensures that persistent players do not lose payoff when the coalition changes. 
Specifically, for each \( i \in \pazocal{S}_{k-1} \cap \pazocal{S}_k \), we require:
\begin{equation}
c_{i,k} \ge \max \left\{
x^{\pazocal{S}_{k-1}}_{i,k} - x^{\pazocal{S}_k}_{i,k}, \; 0
\right\},
\label{eq:comp}
\end{equation}
where \( x^{\pazocal{S}_{k-1}}_{i,k} \) denotes the payoff player \(i\) would receive 
if the previous coalition persisted at epoch $k$.
Positive values of \(x^{\pazocal{S}_{k-1}}_{i,k} - x^{\pazocal{S}_k}_{i,k}\) represent reconfiguration loss. They must be compensated for by \(c_{i,k}\) to preserve participation incentives for players who choose to remain.

The transfers are designed to preserve budget balance across epochs, which requires that total compensations to persistent players are covered by the entry fees ($f_{i,k} \ge 0$) and exit penalties ($p_{i,k} \ge 0$) from joining and departing players, i.e., 
\begin{equation}
\sum_{i \in \pazocal{S}_k \cap \pazocal{S}_{k-1}} c_{i,k}
=
\sum_{i \in \pazocal{S}_k \setminus \pazocal{S}_{k-1}} f_{i,k}
+
\sum_{i \in \pazocal{S}_{k-1} \setminus \pazocal{S}_k} p_{i,k}
\label{eq:balance}
\end{equation}

We impose simple conditions to ensure that entry and exit decisions are individually rational. 
Entrants join only if their payoff after paying the entry fee is non-negative:
\begin{equation}
x^{\pazocal{S}_k}_{i,k} - f_{i,k} \ge 0,
\quad \forall i \in \pazocal{S}_k \setminus \pazocal{S}_{k-1}
\label{eq:fees}
\end{equation}
Similarly, leavers exit only if their payoff after the exit penalty is not positive:
\begin{equation}
x^{\pazocal{S}_{k-1}}_{i,k} - p_{i,k} \le 0,
\quad \forall i \in \pazocal{S}_{k-1} \setminus \pazocal{S}_k
\label{eq:penal}
\end{equation}
Importantly, the payoff \( x_{i,k}^{\pazocal{S}_{k-1}} \) represents the counterfactual payoff that player \( i \) would have received had they remained in the previous coalition. We calibrate the penalty against what the player gives up, not what they would have received under the new coalition from which they are absent.

Given \( \pazocal{S}_{k-1} \), a coalition \( \pazocal{S}_k \subseteq \pazocal{N} \) 
is \emph{feasible} if there exist transfers \( \{f_{i,k}, p_{i,k}, c_{i,k}\} \) 
satisfying~\eqref{eq:comp}--\eqref{eq:penal}.
We denote the set of \emph{feasible} coalitions at epoch \(k\) by
\(
\pazocal{F}_k
\).

Transfer payments and payoff distribution are handled by a trusted third party.
This role is consistent with real-world infrastructure sharing arrangements, 
where centralized entities often manage financial flows and operational coordination~\cite{SECManagementCommittee,FiberCopFY2024, bourreau2020implementing}.

\begin{assumption}
A trusted third party manages the co-investment plan, including revenue collection, payoff distribution, entry fees, and exit penalties. It selects, among feasible coalitions, the coalition that maximizes the total payoff:
\[
\pazocal{S}_k^* \in \operatorname*{argmax}_{\pazocal{S}_k \in \pazocal{F}_k} v_k(\pazocal{S}_k).
\]
\end{assumption}
\subsection{Payoff Allocation}\label{ref:nucleolus}

We allocate the coalition payoff \( v_k(\pazocal{S}_k) \) using the nucleolus, 
a coalitional solution concept that belongs to the core when the core is non-empty. This solution favors stable payoff allocations by minimizing 
the dissatisfaction of coalitions~\cite{schmeidler1969nucleolus}.
For a payoff allocation \(x_k = \{x_{i,k}^{\pazocal{S}_k}\}_{i \in \pazocal{S}_k}\), 
the excess of a sub-coalition \( \pazocal{S}' \subseteq \pazocal{S}_k \) is defined as:
\[
e(\pazocal{S}',x_k) =
v_k(\pazocal{S}') -
\sum_{i \in \pazocal{S}'} x_{i,k}^{\pazocal{S}_k}
\]

The nucleolus is the allocation that lexicographically minimizes the ordered vector 
of coalition excesses, subject to group rationality:
\(
\sum_{i \in \pazocal{S}_k} x_{i,k}^{\pazocal{S}_k}
=
v_k(\pazocal{S}_k).
\)

This allocation minimizes dissatisfaction among the most dissatisfied coalitions, providing a stability-oriented rule for distributing co-investment payoffs.

\subsection{Coalition Formation Algorithm}
At each epoch \( k \), a sequential decision problem must be solved: 
given \( C_{k-1}^* \) and \( \pazocal{S}_{k-1}^* \), we determine the new coalition 
\( \pazocal{S}_k^* \), along with the optimal capacity \( C_k^* \) and allocation $H_k^*$.

Dynamic participation requires enforcing budget balance via transfers. 
We formalize this process through Algorithm~\ref{alg:coalition-selection}.

\begin{algorithm}[H]
\caption{Coalition formation at epoch \( k \)}
\label{alg:coalition-selection}
\begin{small}
\begin{algorithmic}[1]
\Require Player set \( \pazocal{N} \), previous capacity \( C_{k-1}^* \), previous coalition \( \pazocal{S}_{k-1}^* \)
\State Initialize \emph{feasible} coalition set: \( \pazocal{F}_k \gets \emptyset \)

\ForAll{candidate coalitions \( \pazocal{S}_k \subseteq \pazocal{N} \)}
    \State Compute the optimal capacity \( C_k^* \) and value \( v_k(\pazocal S_k) \)\eqref{eq:opt}
    \State Compute payoff allocation \( \{ x_{i,k}^{\pazocal{S}_k} \} \) (e.g., nucleolus)

    \If{CR and GR (Definition~\ref{def:stable}) hold}
        \State Compute transfers \( \{f_{i,k}\}, \{p_{i,k}\}, \{c_{i,k}\} \)~\eqref{eq:fees} \eqref{eq:penal} \eqref{eq:comp}
        \If{budget balance \eqref{eq:balance} holds}
            \State Add \( \pazocal{S}_k \) to \( \pazocal{F}_k \)
        \EndIf
    \EndIf
\EndFor

\State Select optimal coalition:
\(
\pazocal{S}_k^* \in \arg\max_{\pazocal{S}_k \in \pazocal{F}_k} v_k(\pazocal{S}_k)
\)

\State \Return \( \pazocal{S}_k^*, C_k^*, \{f_{i,k}\}, \{p_{i,k}\}, \{c_{i,k}\} \)
\end{algorithmic}
\end{small}
\end{algorithm}

\subsection{Computational Complexity Analysis}

Let \(|\pazocal{N}|\) be the number of players. 
Algorithm~\ref{alg:coalition-selection} explores all candidate coalitions. 
Since coalitions that do not include the NO cannot deploy infrastructure, they can be pruned. 
Thus, only \(2^{|\pazocal{N}|-1}\) coalitions need to be evaluated.
For each candidate coalition \( \pazocal{S}_k \), the capacity and allocation problem~\eqref{eq:opt} is solved once.
The payoff allocation step based on the nucleolus requires considering the excesses of sub-coalitions. 
Since the number of sub-coalitions is exponential in the coalition size, this step has exponential complexity.
Therefore, the overall complexity per epoch is exponential in the number of players.

Since realistic infrastructure co-investment involves a limited number of strategic players, coalition sizes are typically small~\cite{INDERST201428,bourreau2018cooperative,sakr2025co}. 
Hence, the proposed algorithm remains tractable in practice.

\section{Numerical Results}\label{sec:res}

\subsection{Settings}
We fix the parameters and functional forms for the numerical evaluation. These choices illustrate realistic scenarios and may be adapted to specific use cases. Our theoretical analysis remains valid under the assumptions in Section~\ref{sec:model}.

We model SP benefits through a utility function that increases with allocated capacity, 
exhibits diminishing returns, and scales with SP load:
\vspace{-2mm}
\[
u_{i,k}^t(h_{i,k}^t)
=
\beta_i \cdot l_{i,k}^t
\cdot 
\left(1 - 
\exp^{-\xi \, h_{i,k}^t}
\right)
\vspace{-2mm}
\]
where \( \beta_i \) represents the revenue per unit of demand (or request), 
\( l_{i,k}^t \) is the load, i.e., the number of user requests received by SP~\( i \) 
at time slot~\( t \), and \( \xi \) controls how quickly the utility saturates 
as more capacity is allocated. This functional form captures the fact that 
additional capacity initially allows an SP to serve more requests and increase revenue, 
but the marginal benefit decreases once most of the demand can already be handled.

Empirical studies reveal that traffic load follows periodic patterns driven by daily cycles and seasonal trends~\cite{shi2018discovering}. We model it as:
\[
l_{i,k}^t = B_{i,k}^t \cdot S_{i,k}^t \cdot L_{i,k}^t \cdot \delta
\]
where \( B_{i,k}^t \) denotes the request arrival rate (in requests per unit time), \( S_{i,k}^t \) models long-term seasonal trends (e.g., weekly or monthly cycles), \( L_{i,k}^t \) represents the diurnal baseline, and \( \delta \) is the duration of each time slot $t$, so that \( l_{i,k}^t \) is expressed in number of requests.
Following~\cite{sakr2025co}, we define:
\[
L_{i,k}^t = a_0 + \sum_{m=1}^{M} a_m 
\cdot \sin\!\left(\frac{2\pi m (t - t_m)}{24}\right)
\]
where \( a_m \) and \( t_m \) denote amplitude and phase shift, 
chosen differently for each SP to introduce heterogeneity. This makes SPs differ 
not only in their average demand, but also in the timing and intensity of their 
daily traffic peaks.


To reflect practical MEC infrastructure costs, the model captures:
(i) asymmetry between expansion and reduction of capacity,
(ii) reconfiguration overheads, and
(iii) maintenance costs, through:
\vspace{-1mm}
\begin{align}
\mathrm{Cost}(C_k, C_{k-1})
&=
d \,[C_k - C_{k-1}]^+
- \kappa \,[C_{k-1} - C_k]^+
\nonumber
\\
&\quad
+ \gamma\,\mathbf{1}_{\{C_k \neq C_{k-1}\}}
+ d' \,\Delta\, C_k
\label{eq:cost}
\end{align}
Here, $d$ is the unit cost for adding capacity,
$\kappa$ is the monetary income recovered per unit of released capacity,
and $\gamma$ is the fixed cost of technical intervention incurred whenever the capacity changes.
The term $d'\cdot \Delta \cdot C_k$ represents the maintenance cost over the interval duration~$\Delta$,
where $d'$ is the maintenance cost per unit of capacity per unit of time.
\subsection{Evaluation}

Parameters are chosen to reflect realistic MEC co-investment scenarios,
as summarized in Table~\ref{tab:params}.
\begin{table}[h!]
\centering
\caption{Model Parameters}
\label{tab:params}
\renewcommand{\arraystretch}{1.35}
\resizebox{\columnwidth}{!}{%
\begin{tabular}{lll}
\hline
\textbf{Parameter} & \textbf{Value} & \textbf{Description} \\
\hline
$d$        
& \$10.94 per vcore~\cite{azureStackEdgePricing}
& Unit cost for adding capacity \\

$d'$       
& \$0.0225 per hour per vcore~\cite{azureStackEdgePricing}
& Maintenance cost per unit of capacity \\

$\beta_i$  
& \$\(6 \times 10^{-6}\) per request~\cite{AWSLambdaPricing}
& Benefit factor of SP \(i\) \\

$\kappa$   
& \(0.6d\) 
& Income recovered per released capacity \\

$\gamma$         
& \$1000 
& Cost of technical intervention \\

$\xi_i$    
& 0.03 
& Diminishing returns parameter \\

$I$    
& 5 years 
& Investment period \\

$K$    
& 5 epochs 
& Decision epochs \\

$\Delta$   
& 1 year
& Duration of the decision interval \\

$\delta$
& 1 hour
& Duration of each time slot \(t\) \\
\hline
\end{tabular}}
\end{table}

Algorithm~\ref{alg:coalition-selection} takes about 4 minutes per epoch 
for one NO and five SPs on a 13th Gen Intel Core i9-13950HX CPU (2.20~GHz), 
which is practical given the yearly execution of the decision process.

To focus on co-investment scenarios, we consider settings in which the 
NO always participates, reflecting its role as infrastructure owner. 
SP participation varies across epochs depending on the payoff generated 
by the coalition.
\subsubsection{Total Payoff and Capacity}

We consider four co-investment schemes: static, update, dynamic, and update~\&~dynamic.
The static scheme assumes fixed capacity and fixed participation over the entire investment period.
The update scheme allows the infrastructure capacity to be updated across decision intervals while keeping participation fixed.
The dynamic scheme assumes fixed capacity but allows participation to change across epochs.
The update~\&~dynamic scheme combines capacity updates with dynamic participation.

\begin{figure}[h!]
    \centering
    \includegraphics[width=0.8\linewidth]{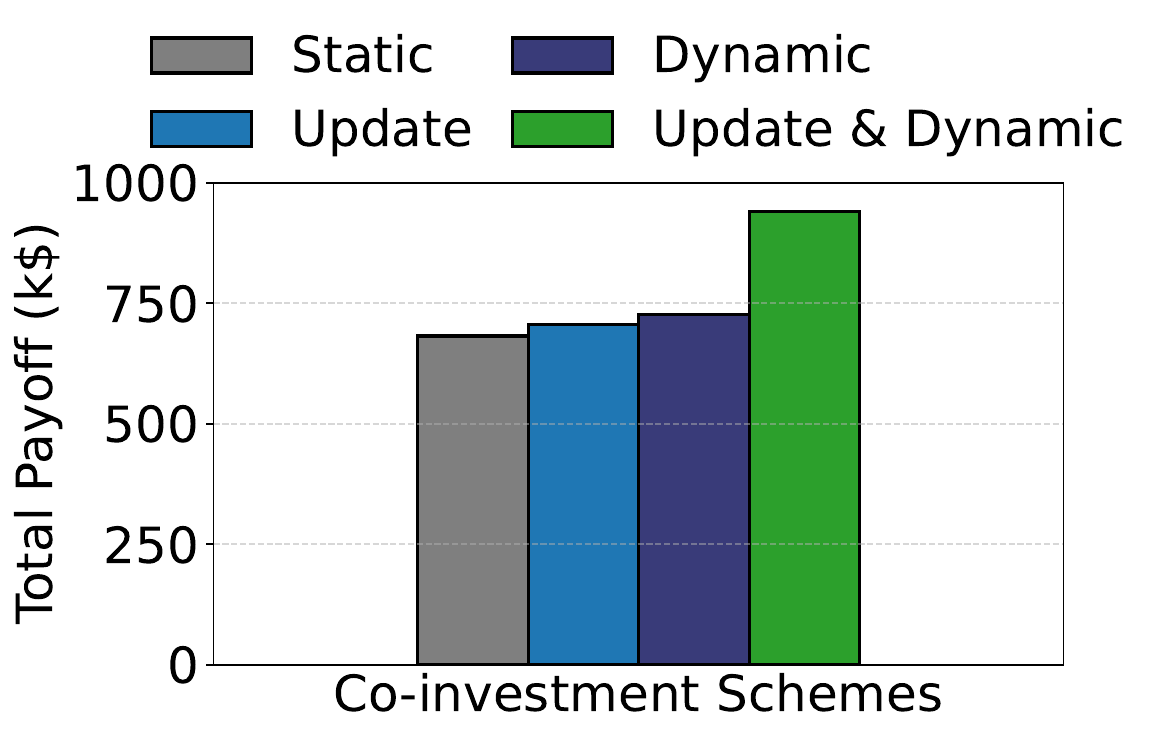}
    \caption{Total payoff across different schemes}
    \label{fig:total}
\end{figure}

\begin{figure}[h!]
    \centering
    \includegraphics[width=0.92\linewidth]{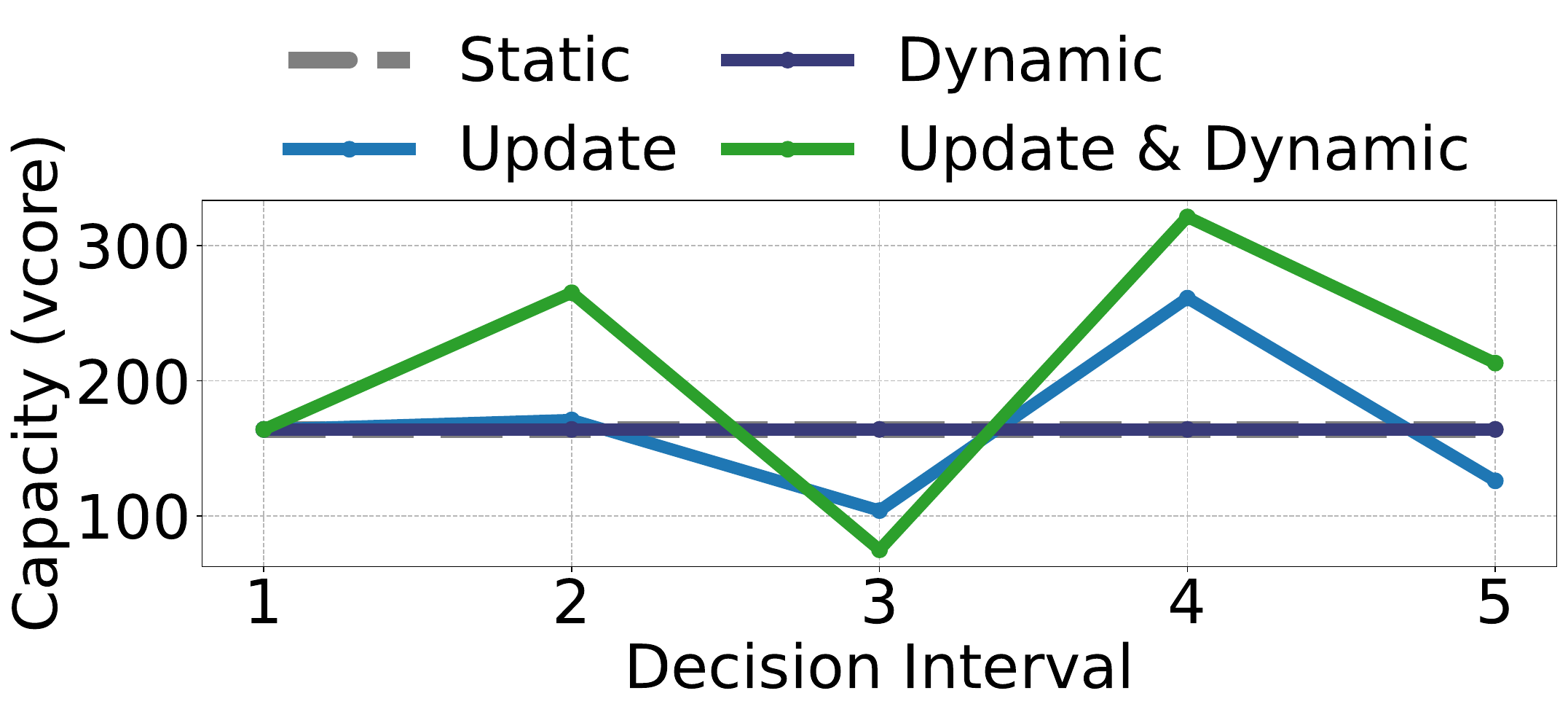}
    \caption{Capacity evolution across intervals}
    \label{fig:capa_traffic}
\end{figure}

\begin{figure}[h!]
    \centering
    \includegraphics[width=0.92\linewidth]{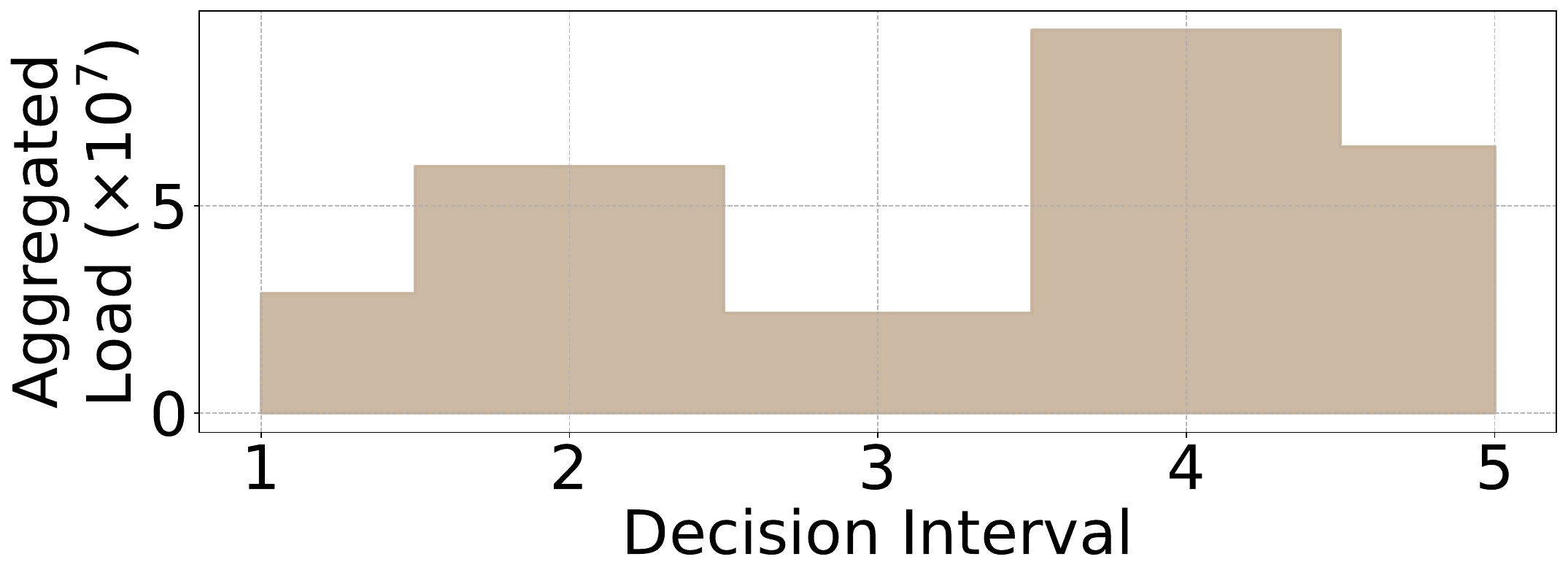}
    \caption{Aggregate end-user demand across intervals}
    \label{fig:traffic}
\end{figure}

\figurename~\ref{fig:total} reports the total payoff (k\$) for the four schemes.
\figurename~\ref{fig:capa_traffic} shows the evolution of infrastructure capacity across decision intervals, while \figurename~\ref{fig:traffic} reports the aggregate end-user demand across decision intervals.

As shown in \figurename~\ref{fig:total}, the update~\&~dynamic scheme achieves the highest total payoff.
This is because it jointly adapts the deployed capacity and the SP participation decisions to the time-varying demand.
When demand is high, such as in interval~4 (see \figurename~\ref{fig:traffic}), updating capacity allows more MEC requests to be served, while dynamic participation enables more SPs to share the investment cost and benefit from the higher revenue opportunity.
When demand is low, such as in interval~3 (see \figurename~\ref{fig:traffic}), capacity can be reduced to avoid unnecessary deployment cost, and participation can also adjust accordingly.
The static scheme obtains the lowest payoff because neither capacity nor participation can respond to demand variations.
The update and dynamic schemes improve performance by adapting only one decision variable, but they remain less effective than the update~\&~dynamic scheme because either participation or capacity is still fixed.

\figurename~\ref{fig:capa_traffic} further explains this result.
The capacity of the update and update~\&~dynamic schemes changes across intervals and follows the aggregate end-user demand trend in \figurename~\ref{fig:traffic}.
Aggregate demand represents the total MEC service requests generated by end users across SPs.
Therefore, higher demand requires more deployed capacity, whereas lower demand requires less capacity.
In contrast, the static and dynamic schemes rely on fixed capacity, which may lead to overprovisioning during low-demand intervals and underprovisioning during high-demand intervals.

Overall, these results show that the best performance is obtained when infrastructure capacity and SP participation are jointly updated, since this better balances served demand, deployment cost, and cost sharing among SPs.

\subsubsection{Network Operator Payoff}

Beyond total coalition payoff, we also evaluate the payoff obtained by the 
NO, since the NO is the infrastructure owner and its incentive 
to participate is essential for MEC deployment.

\begin{figure}[h!]
    \centering
    \includegraphics[width=0.8\linewidth]{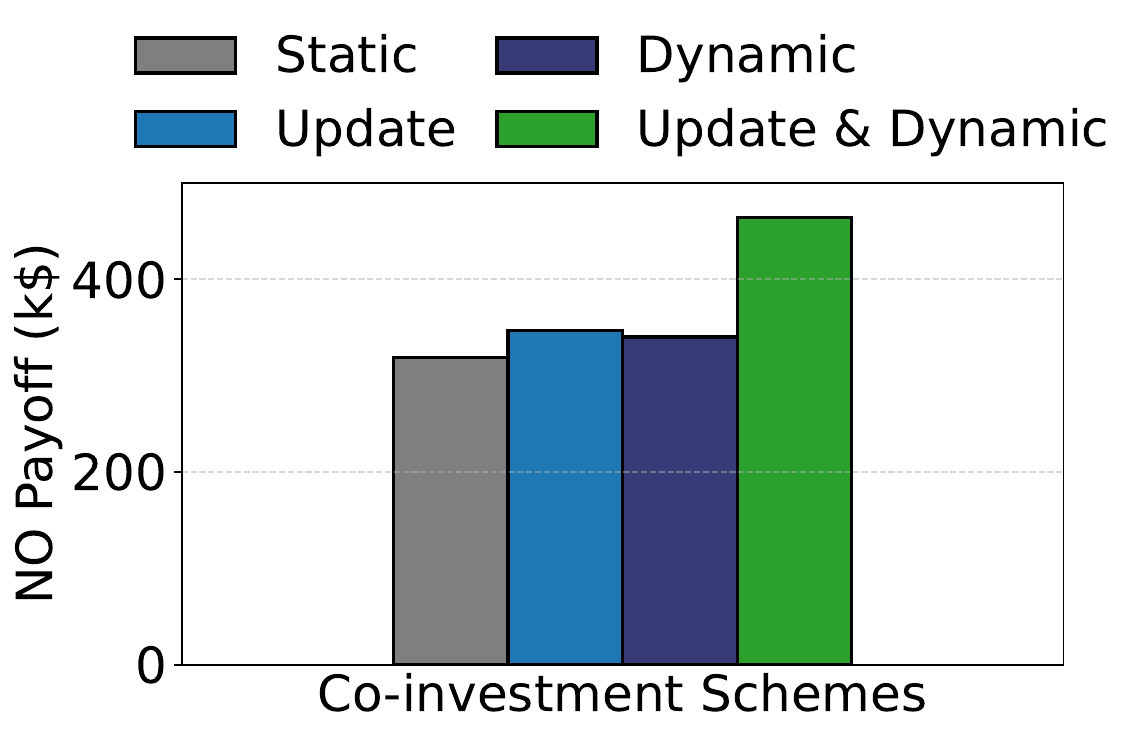}
    \caption{Cumulative payoff of the NO across different schemes}
    \label{fig:no_total_payoff}
\end{figure}

\figurename~\ref{fig:no_total_payoff} reports the cumulative payoff of the NO over the five decision intervals. 
The update~\&~dynamic scheme provides the highest NO payoff, showing that the NO benefits most when both capacity and participation can adapt over time. 
The update scheme outperforms the dynamic scheme because capacity adaptation directly improves infrastructure utilization, which is central to the NO’s role as infrastructure owner. 
By contrast, dynamic participation alone mainly improves the selection of SPs, but cannot correct capacity mismatches. 
The static scheme yields the lowest NO payoff, since it cannot respond to demand or participation changes. 
Overall, these results show that flexibility increases the NO’s incentive to deploy and maintain MEC infrastructure.

\section{Conclusion}\label{sec:conclusion}

In this paper, we modeled a co-investment scheme between a NO and multiple SPs for MEC infrastructure deployment and maintenance as a coalitional game. 
Players can update infrastructure capacity and dynamically join, remain in, or leave the co-investment over time. 
Cooperation is supported through entry fees, exit penalties, and compensations.
Our results highlight the importance of flexibility in co-investment. 
The scheme combining capacity updates and dynamic participation outperforms schemes with fixed capacity and participation, capacity updates only, or dynamic participation only. 
It improves both the total payoff and the NO’s payoff, thereby strengthening its incentive to invest.

Future work includes extending the model to multiple resources and distributed edge nodes, as well as incorporating uncertainty in demand and revenues. 
\section*{Acknowledgment}
This work was supported by the ANR under the France 2030 program, grant NF-NAI: ANR-22-PEFT-0003.


\end{document}